# Collaborative consumption for low and high trust requiring business models: from fare sharing to supporting the elderly and people with disability

## Alex Zarifis*

Karlsruhe Institute for Technology (KIT),
76131 Karlsruhe, Germany
Email: alex.zarifis@kit.edu
Email: alex.e.zarifis@gmail.com
*Corresponding author

## Xusen Cheng

University of International Business and Economics,
10 HuiXin East Street, Beijing, China
Email: xusen.cheng@uibe.edu.cn

## Julia Kroenung

University of Mannheim,
68131 Mannheim, Germany
Email: jkroenun@mail.uni-mannheim.de

**Abstract:** This paper offers an overview of collaborative consumption (CC), the related business models (BM), the value added (VA) from the consumer's perspective and the role of trust. CC is expanding but it is unclear what opportunities it offers and what the challenges will be. This research evaluates the current CC BMs and identifies 13 ways they add value from the consumer's perspective. This research further explores whether CC BMs fall into two categories in terms of what the consumer values. In the first category, the CC BMs require a low level of trust while in the second category of CC BMs a higher level of trust is necessary. It was found that 13 VA by CC BMs could be grouped into personal interest, communal interest and trust building. It is important for organisations to acknowledge how their CC BM relates to these dimensions.

**Biographical notes:** Alex Zarifis has taught and carried out research at the University of Manchester, University of Liverpool, University of Mannheim and Karlsruhe Institute of Technology. He obtained his PhD from the University of Manchester. His research interests are on e-business particularly trust. His research has featured in journals such as *Computers in Human Behavior* and *Information Technology & People*.

Xusen Cheng is the Chair of E-Business at the University of International Business and Economics, Beijing, China. He obtained his PhD from the University of Manchester. His research focuses on e-commerce and the sharing economy. His research has featured in journals such as *Computers in Human Behavior* and *International Journal of Information Management*.

Julia Kroenung is the Chair for E-Business and E-Government at the University of Mannheim. She completed her PhD at the Goethe University, Frankfurt in Germany. Her interests include e-business and social issues in IT. Her research has featured in journals such as *Information & Management*.

# 1   Introduction

Collaborative consumption (CC), where consumers do not purchase a product or service but share it, is growing in popularity (Cusumano, 2018). This is due to a trend away from ownership towards experiencing (Bardhi and Eckhardt, 2012). The terms sharing economy and service sharing are also used to describe this form of exchange (Belk, 2014). This research uses the term CC as it emphasises the collaboration necessary to consume a service or a product in this way. The first two areas of the economy that this business model (BM) disrupted were fare sharing and renting rooms for short periods. Other areas are also influenced but it is unclear which sectors of the economy will be disrupted next.

  CC is at the confluence of some of the biggest trends facing society and consumers. These are moving from ownership to temporary use, social commerce (Puschman and Alt, 2016) climate change, ecological concerns, urbanisation (Cohen and Kietzmann, 2014), demographic change and globalisation. Additionally, local businesses and SMEs are utilising the internet more, workforces are more transient and hobby-jobbing, where a hobby is monetised in a small-scale way, is increasing. Information systems play an important role for all these issues but are some trends where they are the primary disruptor such as digitisation of many aspects of life and mobile commerce (Puschman and Alt, 2016). CC is linked to each of these in both specific practical ways and in the more general prevailing culture and perception at this time. It should be clear how CC has practical implications for climate change with more people situated close to each other in more urban settings. People in their professional and social lives however are also increasingly part of transient teams and fulfil part time roles. They therefore have the necessary skills and are comfortable creating the ad-hoc collaborations involved in sharing products and services. Therefore, the consumer, society and information systems all play an important interrelated role in the ongoing emergence of CC.

  Nevertheless, these new forms of online collaboration are not fully understood. The situation is analogous to the start of the century when innovation in e-commerce gradually gained wide adoption, but the consumer still had strong reservations trusting online consumption. CC still faces a degree of challenge to earn the trust of the consumer (Lee et al., 2018), and the percentage of people that have used CC ranges from 42% in the USA to 1% in Japan (Cusumano, 2018). While there are some high-profile cases of business to consumer CC that have raised the profile of this form of commerce there is also business to business CC which is becoming increasingly popular.

While CC depends on information systems for the collaboration platform it is not necessarily only developments in technology that are decisive in the success or failure of such BMs. The purpose of a BM is to increase the value for all parties involved, so the more value created the stronger the position and the greater the power of the organisations enabling this collaboration (Zott et al., 2011). The success in both fare sharing and room renting was achieved with technology that was available for some years. The challenges and moderators of the success came from other areas such as the legal framework that was not initially compatible. Parallels can be drawn to the early days of e-commerce in the late 90s where to evaluate the possibility of a success lists were created of viable and non-viable BM (Timmers, 1998). A viable BM did not guarantee success, but a non-viable BM almost certainly guaranteed failure. Attempting to identify credible BMs for CC would potentially help in a similar way.

Innovation in BMs is driven by entrepreneurs that want to make a profit and consumers that want a convenient solution that offers the best value. This suggests that smaller niches of the economy, or areas where more public-sector involvement is necessary, such as the elderly and the people with disability (PWD) may not be at the forefront and may be the laggards losing out on possible benefits for years. As in the early days of e-commerce identifying viable BMs, the opportunities and challenges could speed this process up by channelling the limited resources effectively and attracting further resources. The challenges can be identified and overcoming them can be supported. The consumers' 'pull' is as important as the CC organisations' 'push' so avoiding disappointing failures and having more successes will create a positive feedback loop as in transport and accommodation. This research identifies several areas which would benefit from practitioners applying proven or promising CC BMs. Therefore, there are implications for the private and public sector.

The first stage of this research evaluated the prominent CC BMs in terms of their characteristics. The deciding factor in their success was identified as being able to deliver the value the consumer wants more effectively than alternative BMs. Based on the literature, 13 primary ways are identified by which the CC BMs add value from the consumer's perspective. While most research on BM is conceptual, given the pivotal role of the consumer, the second stage of this research utilised empirical data. A survey was carried out that evaluated the 13 value added (VA). The survey evaluated whether there was a difference between CC BMs that required a low or moderate level of trust such as a taxi service and CC BMs that required a high level of trust such as supporting the elderly and PWD. Other possible relationships were also explored including whether these 13 VA fell into the three categories of personal interest, communal interest and trust building. Personal interest added value where the primary perceived benefit was personal while there could be a secondary communal benefit. These include the collaboration platform, lower financial cost, lower effort and enjoyment. Communal interest included those VA where the perceived benefit was primarily communal but could also include some personal interest. These include communal engagement process, utilising existing assets, overcoming barriers and an ethical process. Lastly for trust building VA the benefit was to build the trust necessary for the consumer to make the decision to collaborate. These include reputational trust, trustworthiness of seller based on how they present themselves, institutional trust, information disclosure trust and the concern over the risk of damage.

## 2  CC literature review

### 2.1  CC BM variations

There are currently several variations of CC BMs. They can be summarised in terms of what they offer, whether they moderate or match and what the motivation is to participate as illustrated in Table 1. The table shows how many models share common characteristics between them and can be considered variations on a common theme. In terms of what is offered there are similarities in many of the BM despite them being in different sectors of the economy. Apart from casual carpooling all the other models use an online platform to support the collaboration. There are also some differences, the most important being that some collaborate to share products, some services, some are tangible and some are intangible (Andersson et al., 2013).

A fundamental process in CC is that there is either moderating or matching involved. Moderating can be limited to checking the suitability of content to a more extensive evaluation of the participant, keeping and updating records and resolving conflicts. In most CC BMs the matching process is achieved by information systems which enable the process to be effective and efficient. Motivation is multifaceted in this context and can be divided between the motivation to consume a shared good or service and the motivation to provide them (Bucher et al., 2016). Areas of interest are the propensity of participating in CC, user characteristics and what provides the necessary motivation.

While having a community around an organisation is beneficial to many BMs it is more critical and fundamental here. The community does not just share information and offer motivation but also delivers the product or service. There are additional CC BM models that have a lower level of adoption that are not included in the table such as a food service, an example being Eatwith, tour guides such as Voyable (Ert et al., 2016), neighbourhood local sharing such as TheSharehood (Belk, 2010) and city wide local sharing such as Sharing City Seoul (Cohen and Munoz, 2015). Neighbourhood sharing can be seen as an interesting experiment on how small a CC BM user base can be and exist successfully while city wide sharing can be seen as an experiment in combining CC BMs to create something larger and more comprehensive with synergies.

### 2.2  The VA by CC BMs

Encapsulating a BM has been attempted in several ways with different emphasis and level of abstraction. For a high level of abstraction, the model attempts to succinctly capture what a company does (Timmers, 1998), while more detailed models can represent specific processes. One widely used approach is to identify how value is offered by the model (Zott et al., 2011). Therefore, clarifying the value CC BMs offer can enable a more nuanced understanding of what they are and what distinguishes them from similar concepts, that share some but not all the characteristics. Some of these similar but distinct concepts are co-creation, consuming products from a third party together (Felson and Spaeth, 1978) freemium service (Wagner et al., 2014) and decentralisation, such as decentralised digital currencies (Zarifis et al., 2014).

**Table 1** Components of different CC BMs

| CC BMs | What is offered | Moderating or matching | Motivation | Papers citing the model |
|---|---|---|---|---|
| Sharing of online content (e.g., YouTube) | Provide platform, host content | Moderated by volunteers or professionals | Non-profit, private for profit | Belk (2014) |
| Fare sharing (e.g., Uber) | Matching travellers to part time drivers, no inventory | Matching with system, limited intermediation | Convenient, efficient | Lee et al. (2018) |
| Car rental (e.g., RelayRides) | Individuals rent their cars, no inventory | Matching with system, limited intermediation | Convenience, economical | Sundararajan (2013) |
| Bike sharing (e.g., Nextbike) | Physical objects, service quality, requires inventory | Local government involvement | Profit not primary motivation | Alvarez-Valdes et al. (2014) |
| Ridesharing (e.g., BlaBlaCar) | Provide platform, limited regulation | Matching by system, limited intermediation | Monetary, environmentally friendly | Wagner et al. (2014) |
| Casual carpooling (e.g., San Diego) | No technology mediation, no inventory | Ad-hoc, no centralised organisation | Primarily convenience, monetary saving | Shaheen et al. (2016) |
| Tasks sharing (e.g., TaskRabbit) | Provide platform, identify and vet service providers | Professional moderating and quality controlling | Monetary, convenience | Sundararajan (2013) |
| Space sharing (e.g., Airbnb) | Provide platform, regulate, compensate for damage | Moderates, verifies identities | Convenience, monetary saving, time saving, utilise assets | Hartl et al. (2015) |
| Asset sharing (e.g., Lendogram) | Provide platform, separate into categories, send reminders | Moderating | Monetary, utilise assets | Koopman et al. (2014) |
| Money sharing (e.g., LendingClub) | Evaluate the credit worthiness and create the terms of the loan | Moderating | Monetary, provide loans to people that could not get them | Moehlmann (2015) |
| Investment sharing (e.g., Kickstarter) | Recommend investments | Moderating | Monetary, support innovation | Hamari et al. (2015) |

There are many aspects to CC BMs that are common to other BMs but there are some characteristics, summarised in Table 2, which distinguish them and give them an advantage from the consumer's perspective. While on the one hand looking at the processes within a model is useful, in order to make them more efficient and effective, looking at what the model offers with the eyes of the technology adopter, technology user and the consumer attitude (Kroenung and Eckhardt, 2015) is often decisive. It could be argued that some of the most successful organisations such as Google and Facebook understand the nuances in their consumer and technology user's perspectives better than their competitors. The consumer's perspective is also considered central to CC (Hartl et al., 2015). Some characteristics that are common to many consumption models are expressing identity (Belk, 2014) and service quality that can be considered added value at a given price point (Alvarez-Valdes et al., 2014). For CC one common purpose and way of adding value, of most of these models is efficiency (Sundararajan, 2013). This can be considered as their 'killer application', 'unique selling point' or primary advantage but there are also other significant ways they add value. There are ways of adding value inherent in the nature of most CC and there are other ways of adding value that apply to specific CC BMs. As this research explores the characteristics across sectors, the focus here is on the VA that is inherent in the process of CC and not specific models such as those discussed in the previous section. The VA by the CC BM, from the consumer's perspective can be encapsulated in 13 VA that fall into three categories and are summarised in Table 2. The first category is personal interest and includes the digital collaboration platform, financial cost, effort and enjoyment. The second category has communal benefits and includes utilising assets, overcoming barriers, communal engagement and ethical behaviour. The third category is related to trust building and includes reputational trust, trustworthiness of seller, institutional, information disclosure and risk of damage. These 13 VA and three groups emerged from the literature on CC from the consumer's perspective, but they have parallels to other models of e-commerce that distinguish between the psychological predisposition of the individual, the social influence on the individual and the role of trust. Trust is related to the two preceding factors and to other additional ones (McKnight et al., 2002a). While they are not the same concepts there are similarities between the psychological disposition and personal interest, the social influence and communal interest and the third aspect in both cases is related to trust. This similarity supports the validity of the three categories as CC enabled by information systems can be considered a subsection of e-commerce and is therefore related.

The first VA of personal interest is the collaboration platform VA that includes the information systems the CC BMs provide to the consumer. These systems bring together buyers and sellers; offer a sharing platform and a payment function (Andersson et al., 2013). They are internet-based platforms and their usability and interactivity with mobile devices is often a priority in the design, utilising information such as the location of the user. They offer many benefits, such as enabling access to more options in products or services. Lower cost VA is the lower financial cost for a product or service compared to alternatives that are not CC BMs (Ert et al., 2016). Lower effort VA is the search cost in time and effort, but not monetary, as this is covered by the previous VA. The enjoyment VA covers to what degree the consumer enjoys the activity. Enjoyment is an important factor that has been found to positively influence many actions including selecting one channel to make a purchase over another (Zarifis and Kokkinaki, 2015) and it is also posited to add value in CC.

The first VA of communal interest is related to being part of a community, kind to others (Belk, 2014) and pro-social. It can be considered as non-reciprocal in the specific transaction (Benkler, 2015). Despite this, it may be considered that there will be some reciprocity and some benefit in the future from the sharing community. The collaboration creates a bond between those involved that can be broader and societal (Ert et al., 2016). Communal engagement therefore has both a selfish and selfless element. This is also related to the increasing trend of social shopping (s-shopping) where social networks, enabled by the internet, are utilised by consumers in the process of choosing and purchasing a product in a collaborative way with their friends (Liang and Turban, 2011).

The VA of utilising existing assets of others is related firstly to limiting the sense of waste and secondly the belief of some consumers that used items gain character. It has been argued, that people create narratives for artefacts they have (Chronis, 2015). Therefore, artefacts used in a CC environment may gain a collaborative, green and pro-social narrative which may be appreciated. While people have usually had personal possessions throughout history, economic and technological progress has enabled personal possession to increase. Therefore, consumers have a significant number of possessions that can be considered as stock. Just like companies, they are looking for ways to manage that stock more efficiently. It is not just the information systems or BM enabling CC but also the significant need to offload unused products.

**Table 2**   VA by CC BMs

| | Collaboration platform | Lower cost | Lower effort | Enjoyment | Communal engagement | Utilise assets | Overcome barriers | Ethical | Reputational trust | Trustworthiness of seller | Institutional trust | Info. disclosure trust | Risk of damage |
|---|---|---|---|---|---|---|---|---|---|---|---|---|---|
| Koopan et al. (2014) | x | x | x | | | x | x | | x | | | | |
| Bucher et al. (2016) | x | x | x | | | x | | x | x | | | | x |
| Hamari et al. (2015) | | x | | x | | | | x | x | | x | | |
| Ert et al. (2016) | | x | | | | | | x | x | | | | |
| Puschman and Alt (2016) | | x | x | | | | | | x | x | x | | |
| Kornberger et al. (2018) | x | | x | | x | x | x | x | | | | | |
| Belk (2014) | x | | | x | x | | x | x | x | | x | x | x |
| Andersson et al. (2013) | x | x | x | x | | x | x | | | | | | |

The VA of overcoming 'barriers to entry' refers to making new more efficient solutions possible where this was not allowed in the past due to regulation. In many sectors there are barriers to entry created by regulation. These do not necessarily create an oligopoly, but nevertheless limit competition and can lead to higher prices and lower quality. The obvious example is taxi service. The ethical VA applies to the whole process of purchasing a product or service being moral, green and sustainable. Making the most use of finite resources is increasingly valued in society (Ozanne and Ballantine, 2010).

Trust is usually necessary in transactions and this has been found to be particularly important online (McKnight et al., 2002b). It is also necessary when collaborating online (Cheng et al., 2008) and is hence important for CC. It is considered that the process of sharing in general and CC in particular, requires trust between those participating (Lee et al., 2018). The consumer is burdened with evaluating whether a person or organisation should be trusted and information that allows them to make this evaluation adds value. Therefore, CC BMs add value in terms of trust-building in five ways: Firstly, the CC platform keeps reviews and a record of previous transactions enabling reputational trust (Bucher et al., 2016). The second way that the decision to trust is supported, is based on how the person presents themselves (Ert et al., 2016) and supports their trustworthiness. The seller, or sharer, is allowed to present themselves and the item in question, encouraging trust in themselves. The third way the decision to trust is supported is institutional trust. This is the trust in the various institutions that play a role in ensuring the whole process is fulfilled correctly within the range of what is expected, the regulations and laws. Institutions like the internet, the banking system, regulatory bodies and the government are part of a digitally enabled collaboration and have been found to play a critical role (Zarifis et al., 2014). The fourth aspect of trust building, is the willingness to disclose the personal information necessary for the collaboration and trust that this information will be secure. The fifth and last way the decision on trust is supported is the risk of damage VA. While CC can take many forms, there is usually a risk of damage to those involved in the collaboration. There are usually some safeguards or remedies in place to deal with this. An example is the insurance Airbnb provide its collaborators.

### 2.3  CC for BMs requiring low and high trust

While the literature identifies 13 VA for the consumer from CC BMs, it is unclear whether these VA are equal across different CC BMs. Before answering this question, it needs to be clarified how all the CC BMs can be separated into groups in a meaningful way. If CC BMs can be grouped in a valid and useful way, then the 13 VA can be explored across these groups. It has been shown that perceived risks are negatively correlated with trust, and reduce trusting intention to transact online (Pavlou, 2003). It is posited here that CC BMs can be grouped into low trust requiring BMs and high trust requiring BMs. It is therefore useful to explore whether the 13 VA are different between low and high required trust CC BMs.

For example, one VA the consumer appreciates from CC BMs, is to be convinced to trust that the information they disclose will be secure and not leaked or misused in any way. In many online activities there is a concern about the potential opportunistic behaviour within the context of information privacy (Preibusch et al., 2016) and this includes social networks (Stern and Kumar, 2017). There is concern about potential opportunistic behaviour of online vendors who may sell personal data to third parties such as insurance companies, or allow unauthorised access (Preibusch et al., 2016).

This is also an important dimension from the consumer's perspective in CC where there is a concern about the degree to which their private information is shared. This is a spectrum from complete privacy to sharing all personal information. Firstly, this information sharing can be with the person they are sharing a product or service (Bardhi and Eckhardt, 2012), secondly completely public and open (Bardhi and Eckhardt, 2012), thirdly with the collaboration platform or lastly with the government sector. Trust, particularly reputational trust, has been empirically validated to positively improve the intention to reveal personal information (Eastlick et al., 2006). While personal traits have an impact on trusting beliefs (McKnight et al., 1998), it is useful to explore whether the consumer requires a different level of trust to share their personal information in a relatively low trust requiring situation like fare sharing to a relatively high trust requiring situation like supporting the elderly and PWD. As with the example of the trust related to information disclosure, it is interesting to explore all 13 VA across low and high trust required CC BMs.

Here, the current CC BMs that add value from the consumer's perspective in 13 ways are compared to what CC BMs would have to offer in areas such as supporting the PWD where the VA is perceived differently (Kroenung et al., 2015) and there is a different context in terms of government involvement and regulation. A distinction should be made between CC that requires a relatively low level of trust such as using a fare service with the relatively high required trust of supporting the elderly and PWD. While there are many distinctions that can be made between different CC BMs, as summarised in Table 1, distinguishing based on the level of trust should be explored further as it has not been sufficiently covered. An example of the difference between low and high required trust is the criticism that CC organisations face that they do not cater to the elderly and PWD and do not keep records of how many people from this demographic are served. These concerns extend beyond the period where the service is used, to what the legal and other safeguards are if something goes wrong. In the example of fare sharing, Uber took a number of steps, including strengthening the insurance provided and offering a service for PWD called UberAssist and UberAccess (Ryan, 2016). Firstly, this example shows the importance of offering suitable solutions. This example also shows that successful CC BMs, still have limitations in situations that require a higher level of trust.

For this research, the characteristics of low required trust are to collaborate to access and use a product or service that is perceived by the consumer to be simple, require limited research, the process has limited risk and requires limited trust. Examples are fare sharing and asset sharing such as when you use a collaborator's sports equipment. Characteristics of a high level of trust required are accessing a more complex product or service where more research may be needed to cover the multiple aspects such as regulatory compliance and insurance. There is also a higher risk and need for trust. Examples are supporting the elderly in their daily needs or accessing specialised equipment for the PWD. Part of the difference, is distinguishing between simple and complex products which has been used to better understand user behaviour online (Zarifis and Kokkinaki, 2015). Therefore, the following hypotheses are made based on the 13 VA. The first four hypotheses compare low and high trust in relation to personal VA:

H1   The consumer will appreciate the VA by the collaboration platform equally for CC BMs that require a low and high degree of trust.

H2    The consumer will appreciate the VA by lower cost equally for CC BMs that require a high degree of trust to those that require a low degree of trust.

H3    The consumer will appreciate the VA by lower effort equally for CC BMs that require a high degree of trust to those that require a low degree of trust.

H4    The consumer will appreciate the VA by enjoyment equally for CC BMs that require a high degree of trust to those that require a low degree of trust.

The second four hypotheses compare low and high trust in relation to communal VA:

H5    The consumer will appreciate the VA by the communal engagement process equally for CC BMs that require a high degree of trust to those that require a low degree of trust.

H6    The consumer will appreciate the VA by utilising other's assets equally for CC BMs that require a low and high degree of trust.

H7    The consumer will appreciate the VA by overcoming barriers equally for CC BMs that require a low and high degree of trust.

H8    The consumer will appreciate the VA by ethical process equally for CC BMs that require a high degree of trust to those that require a low degree of trust.

The last five hypotheses compare low and high trust in relation to different aspects of trust building VA:

H9    The consumer will appreciate the VA by reputational trust building more for CC BMs that require a high degree of trust to those that require a low degree of trust.

H10   The consumer will appreciate the VA by building the trustworthiness of seller based on how they present themselves more for CC BMs that require a high degree of trust to those that require a low degree of trust.

H11   The consumer will appreciate the VA by institutional trust building more for CC BMs that require a high degree of trust to those that require a low degree of trust.

H12   The consumer will appreciate the VA by information disclosure trust building more for CC BMs that require a high degree of trust to those that require a low degree of trust.

H13   The consumer will appreciate the VA by mitigating risk of damage more for CC BMs that require a high degree of trust to those that require a low degree of trust.

## 3  Methodology

This research was implemented in two stages the first being an extensive literature review and the second involved empirical analysis. The data collection was necessary to evaluate whether the 13 VA and two categories of low and high required trust CC BMs indicated by the literature review, were valid. The quantitative methodology involved a survey where participants would evaluate the 13 VA. A seven-point Likert scale was used. The criteria for the participant sample was that they had used a form of CC. 461 surveys were completed. Due to an unsatisfactory answer to the question on experience with CC, incomplete surveys and surveys completed too fast, the valid participation was 426.

A paired sample t-test was used to evaluate the hypotheses and whether the difference in the responses to the two CC BM groups were sufficiently different to be significant. Paired sample t-tests are considered suitable for measuring the same person's beliefs in relation to two questions (Pallant, 2007). The responses were screened and cleaned before being analysed. The data was normally distributed and continuous. In order to limit the risk of type 1 errors the p-value for this research was set at 0.05. Descriptive statistics and cluster analysis were used as secondary methods to identify any additional insight and fully utilise the collected data.

## 4 Findings

The findings are summarised in Table 3, illustrated in a spider diagram and scale balance model. Table 3 shows the findings of the t-test that compared the participant's responses to questions on the same VA but across what was posited as low and high required trust CC BMs. The results are shown next to the vignettes used in the survey. The t-test findings are presented in terms of the mean responses to each category, standard error (SE), t-value, p-value in relation to the confidence level of 95% and effect size Pearson's 'r'. The collaboration platform, lower cost and lower effort did not have a significant difference. Enjoyment, communal process, utilising assets and overcoming barriers, ethics, reputational trust, trustworthiness of seller, institutional trust, information disclosure trust and risk had a statistically significant difference between the two groups. The last six also had a strong effect size r and a large difference in the means.

**Table 3** Participant responses for 13 values added across low and high required trust

| | *Value added* | Mean | SE | t(426) | p | r |
|---|---|---|---|---|---|---|
| 1a | The website that I can use to have access to shared transport is valuable to me. | 1.02 | 0.02 | 1.208 | p > 0.05 | 0.06 |
| 1b | The website that I can use to have access to shared care for my elderly relatives is valuable to me. | 0.99 | | | | |
| 2a | When considering using shared transport having a lower financial cost is valuable to me. | 1.78 | 0.03 | 1.80 | p > 0.05 | 0.09 |
| 2b | When considering using shared care for my elderly relatives having a lower financial cost is valuable to me. | 1.73 | | | | |
| 3a | When considering using a shared transport, the whole process needing less effort is valuable to me. | 1.69 | 0.03 | 1.09 | p > 0.05 | 0.05 |
| 3b | When considering using shared care for my elderly relatives the whole process needing less effort is valuable to me. | 1.66 | | | | |
| 4a | When considering using shared transport, the whole process being enjoyable is valuable to me. | 0.54 | 0.02 | 4.18 | p < 0.05 | 0.20 |

**Table 3**   Participant responses for 13 values added across low and high required trust (continued)

| Value added | | Mean | SE | t(426) | p | r |
|---|---|---|---|---|---|---|
| 4b | When considering using shared care for my elderly relatives, the whole process being enjoyable is valuable to me. | 0.46 | 0.02 | 4.18 | p < 0.05 | 0.20 |
| 5a | When considering using shared transport meeting new people and feeling part of the community is valuable to me. | 0.24 | 0.03 | 3.88 | p < 0.05 | 0.19 |
| 5b | When considering using shared care for my elderly relatives meeting new people and feeling part of the community is valuable to me. | 0.11 | | | | |
| 6a | Utilising other's assets is valuable to me when considering using shared transport. | 0.31 | 0.03 | 2.55 | p < 0.05 | 0.12 |
| 6b | Utilising other's assets is valuable to me when considering using shared care for my elderly relatives. | 0.24 | | | | |
| 7a | When considering using shared transport, overcoming barriers, such as the limited competition and availability of taxis, is valuable to me. | 0.01 | 0.02 | 4.84 | p < 0.05 | 0.23 |
| 7b | When considering using shared care for my elderly relatives overcoming barriers, such as the limited competition and availability is valuable to me. | 0.11 | | | | |
| 8a | When considering using shared transport, the whole process being ethical is valuable to me. | 0.65 | 0.04 | 14.67 | p < 0.05 | 0.58 |
| 8b | When considering using shared care for my elderly relatives, the whole process being ethical is valuable to me. | 1.21 | | | | |
| 9a | When considering using shared transport, being able to see evidence of the reputation of the service provider helps me decide who to trust and is valuable to me. | 1.25 | 0.03 | 14.2 | p < 0.05 | 0.57 |
| 9b | When considering using shared care for my elderly relatives being able to see evidence of the reputation of the service provider helps me decide who to trust and is valuable to me. | 1.71 | | | | |
| 10a | When considering using shared transport, how the provider of the service presents themselves online with pictures and text helps me decide who to trust and is valuable to me. | 1.31 | 0.04 | 11.85 | p < 0.05 | 0.50 |
| 10b | When considering using shared care for my elderly relatives how the provider of the service presents themselves online with pictures and text helps me decide who to trust and is valuable to me. | 1.77 | | | | |

**Table 3** Participant responses for 13 values added across low and high required trust (continued)

| | *Value added* | Mean | SE | t(426) | p | r |
|---|---|---|---|---|---|---|
| 11a | When considering using shared transport, the institutions that support and regulate the service help me trust the service and are valuable to me. | 0.83 | 0.04 | 16.7 | p < 0.05 | 0.63 |
| 11b | When considering using shared care for my elderly relatives the institutions that support and regulate the service help me trust the service and are valuable to me. | 1.52 | | | | |
| 12a | When considering using shared transport, feeling secure about disclosing the necessary information is valuable to me. | 1.26 | 0.03 | 17.94 | p < 0.05 | 0.66 |
| 12b | When considering using shared care for my elderly relatives feeling secure about disclosing the necessary information is valuable to me. | 1.87 | | | | |
| 13a | When considering using shared transport, having safeguards in place such as insurance provided by the collaboration platform is valuable to me. | 1.33 | 0.05 | 13.85 | p < 0.05 | 0.56 |
| 13b | When considering using shared care for my elderly relatives overcoming barriers, having safeguards in place such as insurance provided by the collaboration platform is valuable to me. | 1.97 | | | | |

Spider diagrams have been used to summarise large numbers of variables and compare them between two cases in related situations. They have been utilised in the area of information systems and management (Rees-Caldwell and Pinnington, 2013).

The spider diagram in Figure 1 shows the 13 VA, rated across three points on each axis. The three points on each axis present the importance of the VA for the CC BM. While the Likert scale allowed the participant to evaluate each VA from minus three to plus three, the overall averages where all positive. This was expected as the VA were based on the literature. The positive values up to one on the spider diagram can be considered to add moderate VA to CC and are useful. Results from one to one and a half on the spider diagram, can be considered as high VA and important to the CC. Results above one and a half can be considered critical to the CC and without them the CC BM would not be possible. The purpose of the diagram is to offer better granularity between two significantly different iterations of CC BMs. This difference in the value the consumer requires to collaborate in a high level of trust BM will encourage organisations to deliver this value. As the diagram illustrates significant differences, organisations that will want to operate CC BMs with the high trust requirement will have to significantly adapt existing CC BMs that were designed for the low required trust.

The results support three important findings: Firstly, the importance and role of 6 of the 13 VA is very different between high and low trust CC BMs supporting the importance of the distinction between the two groups of BMs. Secondly 5 of the 6 VA with the most significant change are related to trust building. This supports the validity of a difference between low required trust and high required trust CC BMs. The third finding is that beyond the specific ways CC BM models must adapt for the high trust requiring BM, the number of values, for 6 out of 13, and the magnitude of the change, in some cases 32% increase and 21% increase overall illustrate the significant increase in what CC BMs requiring high trust necessitate.

**Figure 1**     Comparative spider diagram of VA by CC BM for low and high required trust (see online version for colours)

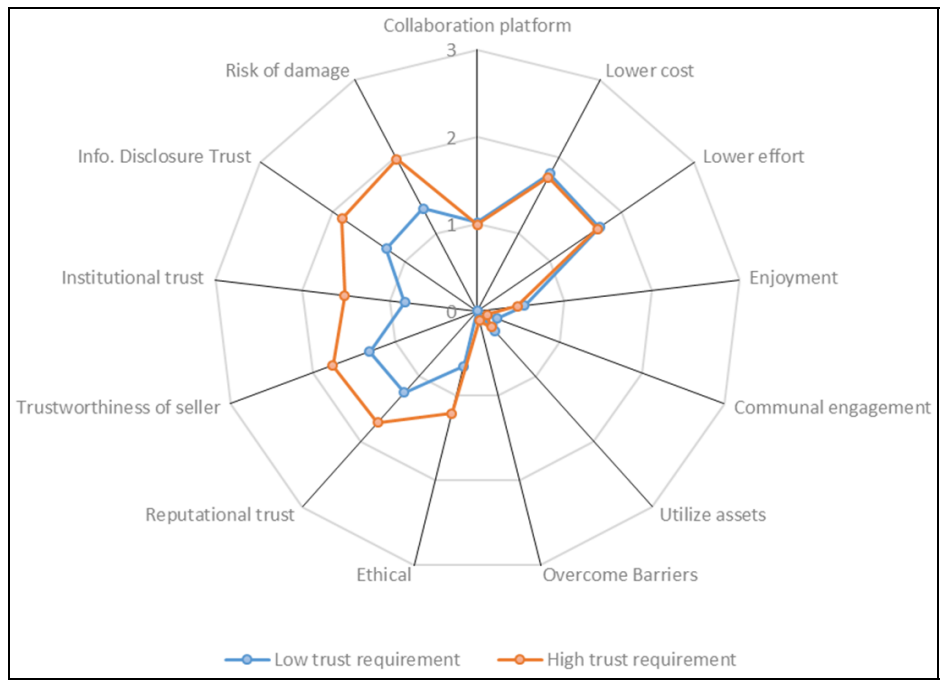

**Table 4**     The findings grouped by personal, communal and trust building VA

|  | *Personal interest* | *Communal interest* | *Trust building* |
| --- | --- | --- | --- |
| Low required trust CC BMs | 1.26 | 0.30 | 1.20 |
| High required trust CC BMs | 1.21 | 0.42 | 1.77 |

Scale balance models are useful for illustrating how the relationship of two variables changes over time, or in two different contexts. The scale balance model in Figure 2 groups VA by personal interest, communal interest and trust building showing their average value as they are presented in Table 4. This is done for low and high required trust CC BMs. The resulting scale balances illustrate how the value offered by the personal VA (P) outweigh trust (T) for the low trust required CC BMs. This relationship is reversed for the high required trust CC BMs where trust building outweighs personal VA. This means that organisations need to make a step change in the way they build trust and that this fundamental trust 'step' must be done at BM level. The scale balances related to communal interest VA (C) are constant, as these VA are less important than personal and trust in all situations.

Beyond the insight gained from the average response to the three categories of VA that were personal, communal and trust it is also of interest that the responses showed more consistency for personal interest and trust building with the communal interest having a broader range of responses. This suggests there is a degree of consensus around personal interest and trust building VA required by the consumer. While ethics VA was placed in the communal group because ethical behaviour from those involved in the collaboration brings both personal and communal benefits, based on the empirical findings it could be argued that it fits better in the trust related VA. This could also be supported theoretically as ethical behaviour has been found to reinforce trust (Leonidou, 2013).

**Figure 2** Difference in VA by personal (P), communal (C) and trust (T) related for low and high trust CC BMs

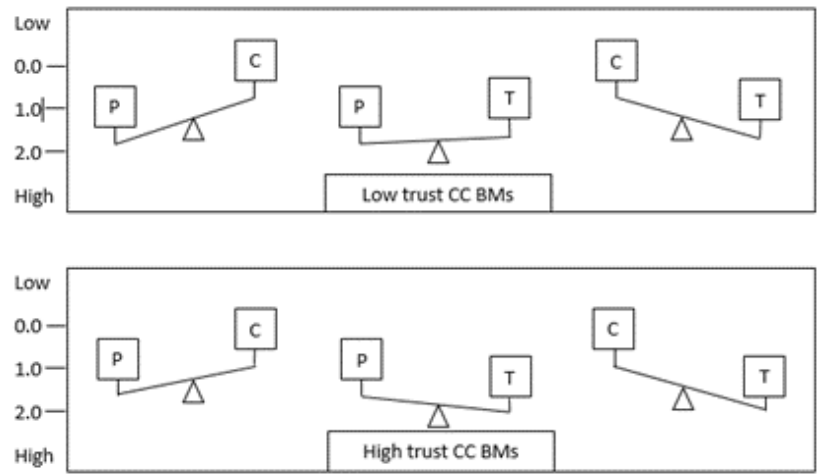

The cluster analysis provided two useful findings. The first finding was related to the communal value average which was low, close to zero. This low average would suggest this VA was not important to the participants. The cluster analysis however identifies two clusters, one cluster where the communal aspect was appreciated across the four related VA and one cluster where it was not appreciated. Therefore, unlike the personal and trust related VA that were valued by all, the communal aspect was valued by a segment of the participants. The second finding of the cluster analysis was that those that valued the enjoyment VA also often valued the communal engagement VA. It can be considered that both of these VA fall under the experience of purchasing or sharing. While for some the experience was important in a broad sense, including enjoyment and communal engagement, others prioritised VA that are related to efficiency.

## 5 Conclusions and implications

### 5.1 Discussion

Some regulation can constrain competition on price and quality (Koopman et al., 2014) by making it harder for new organisations to enter the market and create competition. Information systems and competition brought by CC make some regulations unnecessary and government's strategies and policies must adapt (Baur, 2017). While regulation is a barrier which must be removed for many CC models there are cases where the regulation cannot and must not be removed due to the sensitivity of the service and the necessity of a consistent high quality of service. Some such cases are supporting the elderly and PWD. In this case it is the ability of a BM to encompass regulation rather than remove it that is necessary. Encompassing complex regulations brings challenges for lean models whose purpose up until now has been efficiency. There are examples such as Uber and Airbnb where the original model, that was as simple as possible, was later made more complex by adding more thorough identification and insurance to meet the regulations set by governments but also build consumer trust.

Attracting new stakeholders to CC BMs is important for their development but faces its own challenges. As discussed progress in the CC and the service economy is driven by entrepreneurs and investors who seek out the best return on their investments with simple efficient models. Some more complex or challenging opportunities or niches such as the elderly and PWD may not be initially appealing. If there is a viable model in this area where the government sector invests significant funds and time, investment could be attracted from the private sector or incentives could be provided to the private sector. This could happen in a similar way to how momentum was created for renewable energy in many countries such as the UK and Germany. CC BMs have been shown to fulfil one of the roles of governments, that of coordinating people (Kornberger et al., 2018). They have also been shown to provide synergies when combined in larger scales such as sharing cities (Cohen and Munoz, 2015). In addition to adapting to high trust requiring CC BMs, the future potential can be further developed by merging them with other important disruptive innovations such as the internet of things (IoT) and new renewable powered sources (Kriston et al., 2010). Fully utilising CC BMs across more sectors of the economy and combining disruptive innovations where possible, would compound the efficiencies and synergies they bring.

*5.2 Practical implications*

The findings of this study support the development of CC BMs requiring a high-level trust. We have seen that to better understand and fully utilise CC BMs what the consumer wants from them in terms of VA must be carefully evaluated. By understanding what VA the consumer wants in low and high required trust, the consumer, the organisations involved and society can benefit further. This by-product can be immediate or after some time in the adoption cycle of the innovation. It is therefore pertinent to seek out ways to bring the societal benefits forward. There is however limited research on the benefits of CC BM related to public policy and on collaborating between the government and private sectors (Cohen and Kietzmann, 2014; Kornberger et al., 2018). This private and public collaboration in areas such as supporting the elderly and PWD would fall into the high trust requiring BMs. Building trust in these models is a multifaceted issue that changes in significance depending on the CC BM and the risk to the consumer. The CC BMs that require a high level of trust could in some cases benefit from more involvement and collaboration from the public sector. The public-sector collaboration and government regulation can enhance trust particularly institutional trust and information privacy. It may be that areas such as the elderly and PWD, where people are used to public sector involvement will expect this to satisfy their need for higher trust. The need to build trust applies to all the aspects of institutional trust including the organisations that are involved in ensuring that every aspect of the purchase such as the internet, payment system, regulation and laws about returns.

*5.3 Theoretical implications*

It was identified that the consumer requires 13 VA from the CC BM which can be separated into three categories which are personal interest, communal interest and trust building. The personal interest VA are the collaboration platform, lower cost, lower effort and enjoyment. The communal interest group includes the communal engagement process, utilising assets, overcoming barriers and ethics. The trust building group includes reputational trust, trustworthiness of seller based on how they present themselves, institutional trust, information disclosure trust and managing risk. It was found that CC BMs can be separated into those that require a relatively low level of trust such as fare sharing and those that require a high level of trust such as supporting the elderly and PWD. For the low trust CC BMs the consumer considered the personal interest VA higher while in high required trust CC BMs the consumer valued trust building VA higher. In both cases the communal interest was low apart from ethical VA which was higher for high required trust CC BMs. The findings suggest the change from low requiring trust CC BMs to high requiring trust CC BMs, necessitates a significant improvement in how the organisation builds trust. This can be considered a 'step' change in trust-building which would have to be a consideration at BM level. Iterative improvements at operational level may not increase trust sufficiently.

By gaining insight into the CC BMs organisations use and the VA consumers require, insight was gained on the present and future of this important sector of the economy (Zott et al., 2011). The contribution of this research sheds light on where we are now in CC BMs and supports progress in this important area.

*5.4 Limitations and future research*

There are several limitations. Firstly, the sample was collected from Germany. Other countries may have some variation. Future research can address these limitations by evaluating these issues in other countries. Secondly, while this study sheds light on the nature of trust in CC where higher risk is involved alternative theoretical frameworks should also be explored.

**References**

Alvarez-Valdes, R. et al. (2014) 'Optimizing the level of service quality of a bike-sharing system', *Omega*, Vol. 62, pp.163–175, UK [online] http://dx.doi.org/10.1016/j.omega.2015.09.007.

Andersson, M., Hjalmarsson, A. and Avital, M. (2013) 'Peer-to-peer service sharing platforms: driving share and share alike on a mass-scale', in *International Conference on Information Systems*, pp.1–15.

Bardhi, F. and Eckhardt, G.M. (2012) 'Access-based consumption: the case of car sharing', *Journal of Consumer Research*, December, Vol. 39, pp.881–898.

Baur, A.W. (2017) 'Harnessing the social web to enhance insights into people's opinions in business, government and public administration', *Information Systems Frontiers*, Vol. 19, pp.231–251.

Belk, R. (2010) 'Sharing', *Journal of Consumer Research*, Vol. 36, No. 5, pp.715–734 [online] http://jcr.oxfordjournals.org/lookup/doi/10.1086/612649.

Belk, R. (2014) 'You are what you can access: sharing and collaborative consumption online', *Journal of Business Research*, Vol. 67, No. 8, pp.1595–1600 [online] http://dx.doi.org/10.1016/j.jbusres.2013.10.001.

Benkler, Y. (2015) 'Essay sharing nicely: on shareable goods and the emergence of sharing as a modality of economic production', *The Yale Law Journal*, Vol. 114, No. 2, pp.273–358.

Bucher, E., Fieseler, C. and Lutz, C. (2016) 'What's mine is yours (for a nominal fee) – exploring the spectrum of utilitarian to altruistic motives for internet-mediated sharing', *Computers in Human Behavior*, Vol. 62, pp.316–326 [online] http://linkinghub.elsevier.com/retrieve/pii/S0747563216302679.

Cheng, X., Macaulay, L. and Zarifis, A. (2008) 'Individual trust development in computer mediated teamwork', *WebScience '08: Proceedings of the Hypertext 2008 Workshop on Collaboration and Collective Intelligence*, pp.1–5 [online] http://portal.acm.org/citation.cfm?id=1379157.1379159.

Chronis, A. (2015) 'Substantiating Byzantium: the role of artifacts in the co-construction of narratives', *Journal of Consumer Behaviour*, Vol. 14, pp.180–192 [online] http://oro.open.ac.uk/26883/.

Cohen, B. and Kietzmann, J. (2014) 'Ride on! Mobility business models for the sharing economy', *Organization & Environment*, Vol. 27, No. 3, pp.279–296 [online] http://oae.sagepub.com/cgi/doi/10.1177/1086026614546199.

Cohen, B. and Munoz, P. (2015) 'Sharing cities and sustainable consumption and production: towards an integrated framework', *Journal of Cleaner Production*, pp.1–11.

Cusumano, M.A. (2018) 'Technology strategy and management the sharing economy meets reality', *Communications of the ACM*, No. 1, pp.26–29.

Eastlick, M.A., Lotz, S.L. and Warrington, P. (2006) 'Understanding online B-to-C relationships: an integrated model of privacy concerns, trust, and commitment', *Journal of Business Research*, Vol. 59, No. 8, pp.877–886.

Ert, E., Fleischer, A. and Magen, N. (2016) 'Trust and reputation in the sharing economy: the role of personal photos in Airbnb', *Tourism Management*, Vol. 55, pp.62–73 [online] http://dx.doi.org/10.1016/j.tourman.2016.01.013.

Felson, M. and Spaeth, J.L. (1978) 'Community structure and collaborative consumption – routine activity approach', *American Behavioral Scientist*, Vol. 21, No. 4, pp.614–624.

Hamari, J., Sjöklint, M. and Ukkonen, A. (2015) 'The sharing economy: why people participate in collaborative consumption', *Journal of the Association for Information Science and Technology*, Vol. 14, No. 4, pp.90–103.

Hartl, B., Hofmann, E. and Kirchler, E. (2015) 'Do we need rules for 'what's mine is yours'? Governance in collaborative consumption communities', *Journal of Business Research*, Vol. 69, No. 8, pp.1–8 [online] http://www.sciencedirect.com/science/article/pii/S0148296315006050.

Koopman, C., Mitchell, M. and Thierer, A. (2014) 'The sharing economy and consumer protection regulation: the case for policy change', *The Journal of Business, Entrepreneurship & The Law*, Vol. 8, No. 2, pp.529–545 [online] http://papers.ssrn.com/sol3/papers.cfm?abstract_id=2535345.

Kornberger, M. et al. (2018) 'Rethinking the sharing economy: the nature and organization of sharing in the 2015 refugee crisis', *Academy of Management Discoveries*, Vol. 4, No. 3, pp.314–335.

Kriston, A., Szabo, T. and Inzelt, G. (2010) 'The marriage of car sharing and hydrogen economy: a possible solution to the main problems of urban living', *International Journal of Hydrogen Energy*, Vol. 35, No. 23, pp.12697–12708.

Kroenung, J. and Eckhardt, A. (2015) 'The attitude cube – a three-dimensional model of situational factors in IS adoption and their impact on the attitude-behavior relationship', *Information and Management*, Vol. 52, No. 6, pp.611–627 [online] http://dx.doi.org/10.1016/j.im.2015.05.002.

Kroenung, J., Jaeger, L. and Kupetz, A. (2015) 'System characteristics or user purpose? – A multi-group analysis of online shopping by mobility impaired and unimpaired users', *23rd European Conference on Information Systems (ECIS)*, pp.1–17.

Lee, Z.W.Y. et al. (2018) 'Why people participate in the sharing economy: an empirical investigation of Uber', *Internet Research*, Vol. 28, No. 3, pp.829–850.

Leonidou, L. et al. (2013) 'Business unethicality as an impediment to consumer trust: the moderating role of demographic and cultural characteristics', *Journal of Business Ethics*, Vol. 112, No. 3, pp.397–415.

Liang, T. and Turban, E. (2011) 'Introduction to the special issue social commerce: a research framework for social commerce', *International Journal of Electronic Commerce*, Vol. 16, No. 2, pp.5–14 [online] http://www.tandfonline.com/doi/full/10.2753/JEC1086-4415160201.

McKnight, H., Choudhury, V. and Kacmar, C. (2002a) 'Developing and validating trust measures for e-commerce: an integrative typology', *Information Systems Research*, Vol. 13, No. 3, pp.334–359.

McKnight, H., Choudhury, V. and Kacmar, C. (2002b) 'The impact of initial consumer trust on intentions to transact with a web site: a trust building model', *Journal of Strategic Information Systems*, Vol. 11, pp.297–323.

McKnight, H., Cummings, L. and Chervany, D. (1998) 'Initial trust formation in new organizational relationships', *Academy of Management Review*, Vol. 23, No. 3, pp.473–490.

Moehlmann, M. (2015) 'Collaborative consumption: determinants of satisfaction and the likelihood of using a sharing economy option again', *Journal of Consumer Behaviour*, Vol. 14, pp.193–207.

Ozanne, L. and Ballantine, P. (2010) 'Sharing as a form of anti-consumption? An examination of toy library users', *Journal of Consumer Behaviour*, Vol. 9, No. 6, pp.485–498.

Pallant, J. (2007) *SPSS Survival Manual*, 3rd ed., McGraw Hill, Maidenhead.

Pavlou, P.A. (2003) 'Consumer acceptance of electronic commerce: integrating trust and risk with the technology acceptance model', *International Journal of Electronic Commerce*, Vol. 7, No. 3, pp.69–103.

Preibusch, S. et al. (2016) 'Shopping for privacy: purchase details leaked to PayPal', *Electronic Commerce Research and Applications*, Vol. 15, pp.52–64.

Puschman, T. and Alt, R. (2016) 'Sharing economy', *Business & Information Systems Engineering*, January, Vol. 58, pp.93–99.

Rees-Caldwell, K. and Pinnington, A.H. (2013) 'National culture differences in project management: comparing British and Arab project managers' perceptions of different planning areas', *International Journal of Project Management*, Vol. 31, No. 2, pp.212–227 [online] http://dx.doi.org/10.1016/j.ijproman.2012.04.003.

Ryan, L. (2016) 'Digital disruption in the NFP sector', *Company Director*, Vol. 32, No. 3, pp.14–15 [online] http://search.informit.com.au/documentSummary;dn=188433682559773;res=IELAPA.

Shaheen, S.A., Chan, N.D. and Gaynor, T. (2016) 'Casual carpooling in the San Francisco Bay Area: understanding user characteristics, behaviors, and motivations', *Transport Policy*, pp.1–9 [online] http://linkinghub.elsevier.com/retrieve/pii/S0967070X16300038.

Stern, T. and Kumar, N. (2017) 'Examining privacy settings on online social networks: a protection motivation perspective', *International Journal of Electronic Business*, Vol. 13, Nos. 2/3, pp.244–272 [online] http://www.inderscience.com/info/inarticle.php?artid=83327.

Sundararajan, A. (2013) 'From Zipcar to the sharing economy', *Harvard Business Review*, pp.4–7 [online] http://blogs.hbr.org/cs/2013/01/from_zipcar_to_the_sharing_eco.html.

20     *A. Zarifis et al.*

Timmers, P. (1998) 'Business models for electronic markets', *Electronic Markets*, Vol. 8, No. 2, pp.3–8.

Wagner, T.M., Benlian, A. and Hess, T. (2014) 'Converting freemium customers from free to premium: the role of the perceived premium fit in the case of music as a service', *Electronic Markets*, Vol. 24, No. 4, pp.259–268.

Zarifis, A. and Kokkinaki, A. (2015) 'The relative advantage of collaborative virtual environments and two-dimensional websites in multichannel retail', in *Lecture Notes in Business Information Processing*, pp.233–244 [online] http://link.springer.com/10.1007/978-3-319-19027-3_19.

Zarifis, A. et al. (2014) 'Consumer trust in digital currency enabled transactions', *Lecture Notes in Business Information Processing*, Vol. 183, pp.241–254.

Zott, C., Amit, R. and Massa, L. (2011) 'The business model: recent developments and future research', *Journal of Management*, Vol. 37, No. 4, pp.1019–1042.